\documentclass[aps,twocolumn,showpacs,preprintnumbers,amsmath,amssymb,floatfix]{revtex4-1}

\usepackage{graphicx,tabularx}
\usepackage{dcolumn}
\usepackage{bm}
\usepackage{color}
\usepackage{epstopdf}
\usepackage{upgreek}
\usepackage{color}

\begin{document}


\title{Defect-induced magnetism in TiO$_2$: An example of quasi 2D magnetic order with perpendicular anisotropy}

\author{
Markus Stiller\,$^{\dagger}$ and Pablo
  D. Esquinazi\,$^{*}$}
\affiliation{Division of Quantum Magnetism and
Superconductivity, Felix Bloch Institute for Solid
State Physics, Faculty of Physics and Earth Sciences, University
of Leipzig, Linn\'estrasse 5, D-04103 Leipzig, Germany}

\begin{abstract}

Magnetic order at room temperature induced by atomic lattice defects,
like vacancies, interstitials or pairs of them, has been observed in a
large number of different nonmagnetic hosts, such as pure graphite,
oxides and silicon-based materials. High Curie temperatures and time
independent magnetic response at room temperature indicate the
extraordinary robustness of this new phenomenon in solid state
magnetism. In this work, we review experimental and theoretical
results in pure TiO$_2$ (anatase), which magnetic order can be
triggered by low-energy ion irradiation. In particular, we discuss the
systematic observation of an ultrathin magnetic layer with
perpendicular magnetic anisotropy at the surface of this oxide.\\

\noindent $\star$: esquin@physik.uni-leipzig.de\\
$\dagger$: Current address: zollsoft GmbH, Jena, Germany. Email: markus@mstiller.org \\
  
\end{abstract}

\maketitle

\section{Introduction to Defect-induced Magnetism}
\vspace{-0.5cm}
Till not so far away in time, solid state physicists and materials
scientists were convinced that to get magnetic order in a solid one
needs a certain amount of magnetic ions, like Fe, Ni, etc., in the
atomic lattice. Their amount as well as their environment have a
direct influence in the magnetic ordering temperature, i.e., the Curie
temperature. This concept was and still is successfully applied in
basic and applied research to get magnetic order in solids, since
Heisenberg introduced the idea of exchange interaction between the
electron orbits of neighbor magnetic
ions~\cite{heisenberg28}. Actually, the changes in the electron orbits
produced by a defect, such as a vacancy in its environment, can be
substantially large. Therefore, these changes can lead to a
non-negligible probability to have a significant local magnetic
moment, see
e.g.~\cite{kha09,fis11,lor15apl,esq20,vol10,ogale10,sto10,dim13}.  To
get magnetic order in a solid through atomic lattice defects or
through doping of non magnetic ions we need to reach a minimum defect
density of the order of (or larger than) $\sim 5$ at.\%.  The reason
is that at this or higher density, the exchange coupling mechanism
between the localized magnetic moments at the defects gets robust
enough to trigger the alignment between them.

Defect engineering is also of importance 
in two-dimensional samples, as in transition-metal dichalcogenides
materials, see, e.g., the review in \cite{lia21}. Moreover, magnetic order through 
Se-vacancies
has been recently demonstrated in monolayer VSe$_2$\cite{chu20}. 
Ion irradiation can be used in this case to systematically 
produce a certain kind of vacancy by appropriately choosing the
ion and its energy. 
\section{Emerging ferromagnetic phase through ion irradiation}

Ion irradiation is a sophisticated method to produce atomic lattice
defects systematically at certain positions and a given density. The
main difference between the irradiation of a solid with different
kinds of ions (including protons and electrons) is given by their
penetration depth, density and type of atomic lattice defects produced
upon selected energies.  Several works on the subject were published
in the past, see
e.g.~Refs.~\cite{hong_room-temperature_2006,zhou_origin_2009,cruz_ferromagnetism_2009,thakur_irradiation_2011,yoon_magnetic_2007,li_defect_2013,odin_2021}
with Curie temperatures up to
880~K~\cite{yoon_oxygen-defect-induced_2006,yoon_magnetic_2007}.

The possibility to trigger local magnetic
moments up to magnetic order by particle irradiation on a given solid,
whatever its structure, is significant.  Starting with proton
irradiation of graphite \cite{pabloprl03,ohldagl,ohldagnjp,chap3} and
ZnO~\cite{lor15apl} to Ar$^+$ irradiation of
TiO$_2$~\cite{sti16,bos21}, the amount of published works using
irradiation to trigger magnetic order increases steadily. 

In this section we would like to discuss general results following the
theoretical works described in Refs.~\cite{Stiller2020,bos21}.  In
particular we discuss here the emergence of the two-dimensional
magnetic order at the surface of TiO$_2$ in its anatase structure,
obtained assuming Ar$^+$ irradiation at low energies
$\lesssim 200~$eV. \textcolor{black}{In general, after ion irradiation with the corresponding fluence and energy to
trigger magnetic order, the defect density remains below a threshold where amorphicity grows all over the sample.
The magnetic order is in general observed directly after ion irradiation without any further (thermal) treatment.}

Which are the main magnetic defects one produces in TiO$_2$ by Ar$^+$
irradiation?  From the molecular dynamics simulations of collision
cascades in anatase TiO$_2$ given in~\cite{rob14,Stiller2020,bos21},
the primary magnetic defects are the so-called di-Frenkel pairs (dFP),
consisting of two Ti atoms displaced into interstitial sites leaving
behind two vacancies. Also oxygen vacancies O$_{\rm v}$ are
created. With the help of density functional theory (DFT)
calculations, the magnetic moment of a dFP has been calculated to be
$2~\upmu_B$~\cite{Stiller2020} and $1~\upmu_B$ \citep{Li2007} for the
O$_{\rm v}$.  The defect formation probabilities for these defects,
calculated in~\cite{rob14}, are large ($\sim 40\%$ and $\sim 50\%$,
respectively).  A diagram of these probabilities also for other defects
in anatase TiO$_2$ can be seen in Fig.~8 in~\cite{bos21}).  With the
knowledge of these probabilities and using the program SRIM
\cite{ziegler2} a magnetic phase diagram can be proposed.

\onecolumngrid

\begin{figure}[h!]
\begin{center}
\includegraphics[width=18cm]{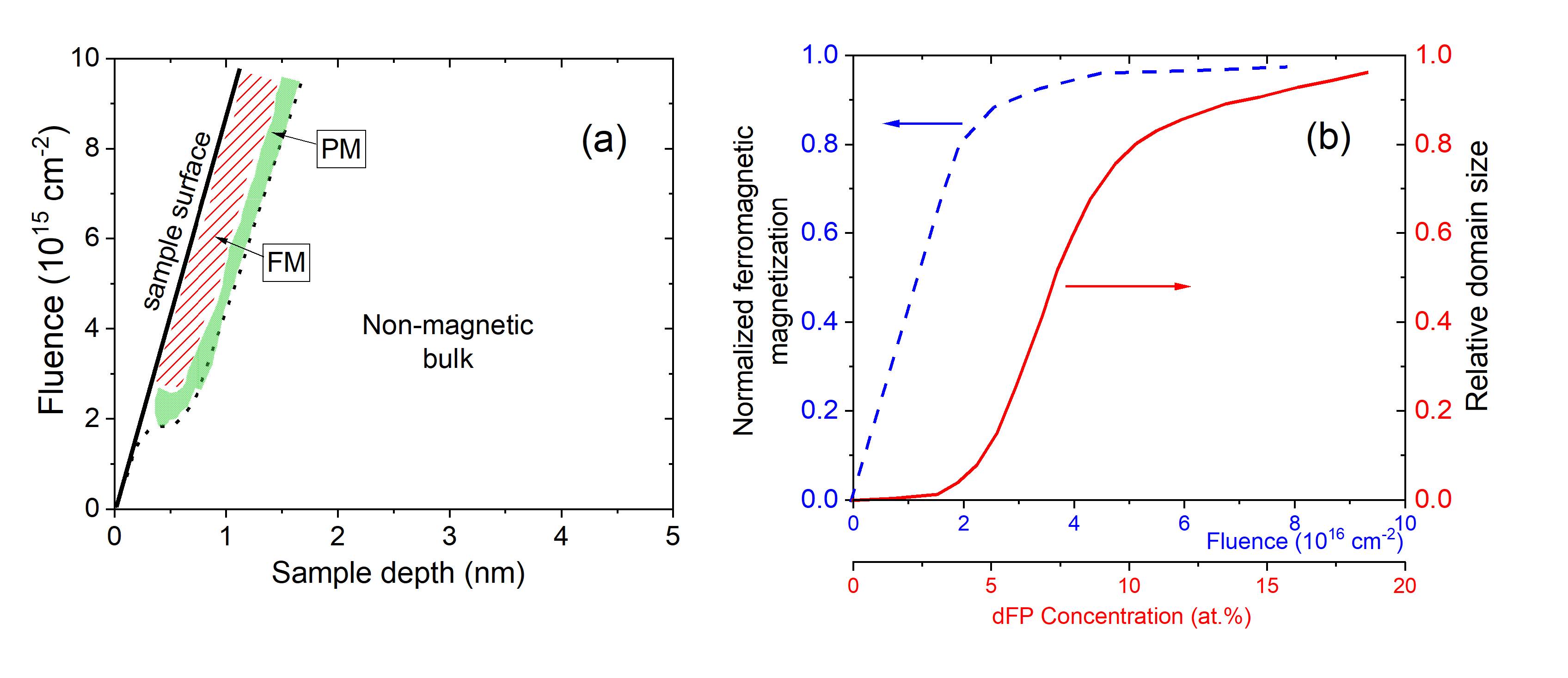}
\end{center}
\vspace{-0.5cm}
\caption{(a) Semiquantitative magnetic phase diagram (fluence
  vs. sample depth) for a TiO$_2$ anatase sample irradiated with
  Ar$^+$ ions of 200~eV energy following the results of
  \cite{bos21}. The straight line denotes the position of the sample
  surface, which shifts due to the sputtering. The red region is the
  ferromagnetic (FM) region and the green a paramagnetic (PM) one. The
  dotted line represents roughly the transition region between a PM
  and a non-magnetic one, where the mean number of defects created by
  the irradiation is negligible. The dotted line can be considered as
  the penetration depth of the irradiated ions. (b) Blue dashed line:
  Normalized ferromagnetic magnetization vs. Ar$^+$ fluence or the
  estimated dFP density in TiO$_2$ anatase phase. Red line: Relative
  FM domain size vs. fluence or dFP concentration. The curves
  represent semiquantitatively the main theoretical results from
  \cite{bos21} for the case of irradiation of Ar$^+$-ions with 200~eV
  energy in TiO$_2$ anatase phase. }\label{Fig-L1}
\end{figure}

\twocolumngrid 

Figure~\ref{Fig-L1}(a) shows diagrammatically the magnetic phases we
expect as a function of the used fluence and sample depth.  The
magnetic phase diagram resumes the results obtained by \cite{bos21}
for TiO$_2$ anatase using Ar$^+$ ions.  The absolute values included
in the magnetic phase diagram of Fig.~\ref{Fig-L1} roughly correspond
to the case we discuss here (Ar$^+$ ions at an energy of
$\sim 200$~eV). The straight line in Fig.~\ref{Fig-L1}(a) represents
the evolution of the surface position of the TiO$_2$ sample (which
represents a decrease of the total sample thickness), when the ion
fluence increases. The main reason for this behavior is surface
sputtering, non negligible at low ion energies. Roughly speaking,
using SRIM \cite{ziegler2} the estimated sputtering is
$\sim 1~$nm/($10^{16}$ ion/cm$^{2}$) \cite{bos21}. Due to the
sputtering the amorphous surface layer produced by the irradiation is
continuously removed. Following \cite{bos21}, from a fluence value
$\sim 4 \times 10^{15}~$cm$^{-2}$ the defect creation and sputtering
processes reach equilibrium where the volume (and defect density) of
the emerging FM phase (red region in Fig.~\ref{Fig-L1}(a)) remains
constant over the whole fluence range above this value.  In this
regime the thickness of the FM phase reaches a value of
$d_{\rm FM} \simeq 0.46$~nm, which is about $1/2$ of the anatase
lattice constant $c = 0.951~$nm along the (001) crystal direction. It
means that with an irradiation energy of $\sim 200$~eV we expect to
have an emerging FM phase at the first $\sim$ two layers of the
TiO$_2$ lattice at its surface.

Deeper in the sample, beyond the $\sim 0.5~$nm thick FM layer, the
density of magnetic defects decreases below the required minimum
($\sim 5\%$) to get FM. Instead, a paramagnetic (PM) phase appears
(green region in Figure~\ref{Fig-L1}(a)). The dotted line in the diagram
represents the transition region between PM and the non-magnetic bulk,
which can be interpreted as the mean penetration depth of the Ar$^+$
ions at 200~eV.

Increasing the defect density we expect a transition from isolated
local magnetic moments to a long-ranged ordered phase.  How much of
the dFP defects are created in the TiO$_2$ anatase atomic lattice
within the FM phase?  This depends on the length-scale of the exchange
coupling, which determines whether two localized magnetic moments (the
dFP in the case of TiO$_2$) are close enough to each other to interact
ferromagnetically.  Ref.~\cite{bos21} answered this question within
the framework of percolation theory and the main results are shown
diagrammatically in Fig.~\ref{Fig-L1}(b). It semiquantitatively shows
the theoretical results \cite{bos21} on the evolution of the FM
magnetization and the relative domain size as a function of dFP
concentration or fluence for the irradiation of Ar$^+$-ions at 200~eV.
Single pairs of defects interacting ferromagnetically build the
smallest ferromagnetic domain. Increasing the defect density the size
of the domains grows as shown by the red line in
Fig.~\ref{Fig-L1}(b). For a density of Frenkel pairs
dFP~$\gtrsim 10$~at.\%, the domain size tends to saturate at the sample
size (relative domain size 1). At this defect concentration the FM
magnetization (blue dashed line in Fig.~\ref{Fig-L1}(b)) tends to
reach its saturation.

 The magnetic percolation transition as a function of the fluence of
 Ar$^+$-ions (or defect density), was experimentally verified by
 measuring the remanent magnetic moment at zero field of TiO$_2$ thin
 films as a function of the fluence.  It follows the expected critical
 behavior for a percolation transition of a magnetic bilayer system
 (see Fig.~13 in~\cite{bos21}). In the next section we review the
 experimental evidence for the appearance of this 2D magnetic system
 with the interesting property of having the magnetization easy-axis
 normal to the main sample surface.

\section{Evidence for  a  two-dimensional ferromagnetic phase with out-of-plane easy axis}

\begin{figure}
\begin{center}
\includegraphics[width=0.9\columnwidth]{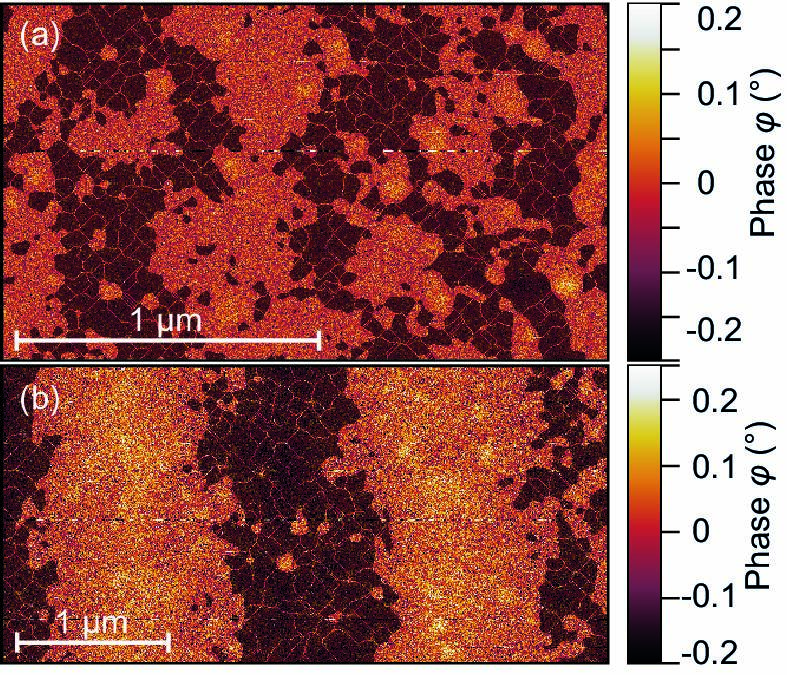}
\end{center}
\caption{
\textcolor{black}{Magnetic force microscopy measurements at two different positions (a) and (b) of a fully
irradiated and otherwise untreated anatase thin film. The TiO$_2$ thin film was irradiated with Ar$^+$-ions
with a fluence $\sim 10^{16}$cm$^{-2}$ and 200~eV energy.
 The magnetic domains have been segmented with a barrier of 40\%
and a Gaussian smoothing factor of 8px.}}
\label{mfm1}
\end{figure}

\subsection{Reasons for the magnetic anisotropy}

As indicated above, following theoretical and experimental results, at
a low ion energy of 200~eV the magnetically ordered phase appears at
the first two layers of the TiO$_2$ anatase surface. If we irradiate
TiO$_2$ with Ar$^+$-ions of higher energy (e.g., 500~eV), the magnetic
anisotropy changes and the magnetization easy-axis points parallel to
the sample surface \cite{Stiller2020,bos21}. This evidence clearly
indicates that the negative magnetic anisotropy energy (MAE) found at
low irradiation energies is directly related to the two-dimensionality
of the ferromagnetic phase produced at the surface of the TiO$_2$
sample.  The localized magnetic moments of the dFP defect at the (001)
anatase surface of the measured samples are at the two interstitials
Ti places. One of them Ti$_{i,2}$ is located at the first surface
layer, whereas the other interstitial Ti$_{i,1}$ on the second
layer. According to DFT electronic structure calculations using the
full potential linearized augmented plane wave (FLAPW) method, the
magnetic defect Ti$_{i,1}$ shows a similar spin structure as in the
bulk. Whereas the Ti$_{i,2}$ defect shows a completely different spin
polarization with a negative MAE due to the reduced coordination of
the surface \cite{bos21}. Increasing the magnetically ordered volume
inside the sample by increasing the irradiation energy, the smaller is
the relative contribution of this magnetic surface to the total
magnetization and the magnetic anisotropy turns to positive.

The first clear hints for the unusual magnetic anisotropy of the
TiO$_2$ thin films after irradiation were obtained by SQUID angle
dependent magnetization measurements~\cite{sti16}. This interesting
finding was supported a few years later through similar SQUID
measurements of new TiO$_2$ thin films irradiated at different
energies~\cite{Stiller2020,bos21}. We would like to emphasize the main
results obtained from magnetic hysteresis loops obtained applying
external magnetic fields parallel and perpendicular to the film
surface of irradiated TiO$_2$ samples using a SQUID magnetometer.  The
total magnetic anisotropy energy was obtained from the difference
between the areas of the two first field (virgin) dependent curves,
for more details see Ref.~\cite{bos21} and its supplementary
information~\cite{si21}.  The magnetization results show that the MAE
for the 200~eV Ar$^+$ irradiated sample is negative with a value of
MAE~$\sim - 0.03~$mJ/cm$^2$ nearly independent of the irradiated
fluence up to $\sim 3 \times 10^{16}~$cm$^{-2}$ \cite{bos21}, in
agreement with the predicted behavior given by the red area in
Fig.~\ref{Fig-L1}(a). In the next section we discuss the visualization
of magnetic domains via MFM, supporting the anomalous MAE of
irradiated TiO$_2$.  \textcolor{black}{It is worth to note that 
perpendicular magnetic anisotropy has been reported 
in monolayers of Cr$_3$Te$_4$,  triggered in this case not by atomic lattice defects 
in the material itself  but by an interfacial effect
between the monolayer and graphite \cite{chu21}.}

\subsection{Magnetic Force Microscopy}

Thin films of anatase, prepared by pulsed laser deposition on
LaAlO$_3$ substrate, have been irradiated with low energy ions, and
measured using magnetic force microscopy
(MFM)~\cite{Stiller2020}. Previously conducted SQUID measurements show
a low remanence, thus we expect randomly distributed magnetic domains
at the anatase surface. Figures~\ref{mfm1}(a) and (b) show MFM
measurements at two different positions, and the magnetic domains as
well as their out-of-plane character can clearly be recognized. An
in-plane domain structure would only be visible at the domain walls as
the out-of-plane field vanishes within the domains.
\begin{figure}
\begin{center}
\includegraphics[width=0.9\columnwidth]{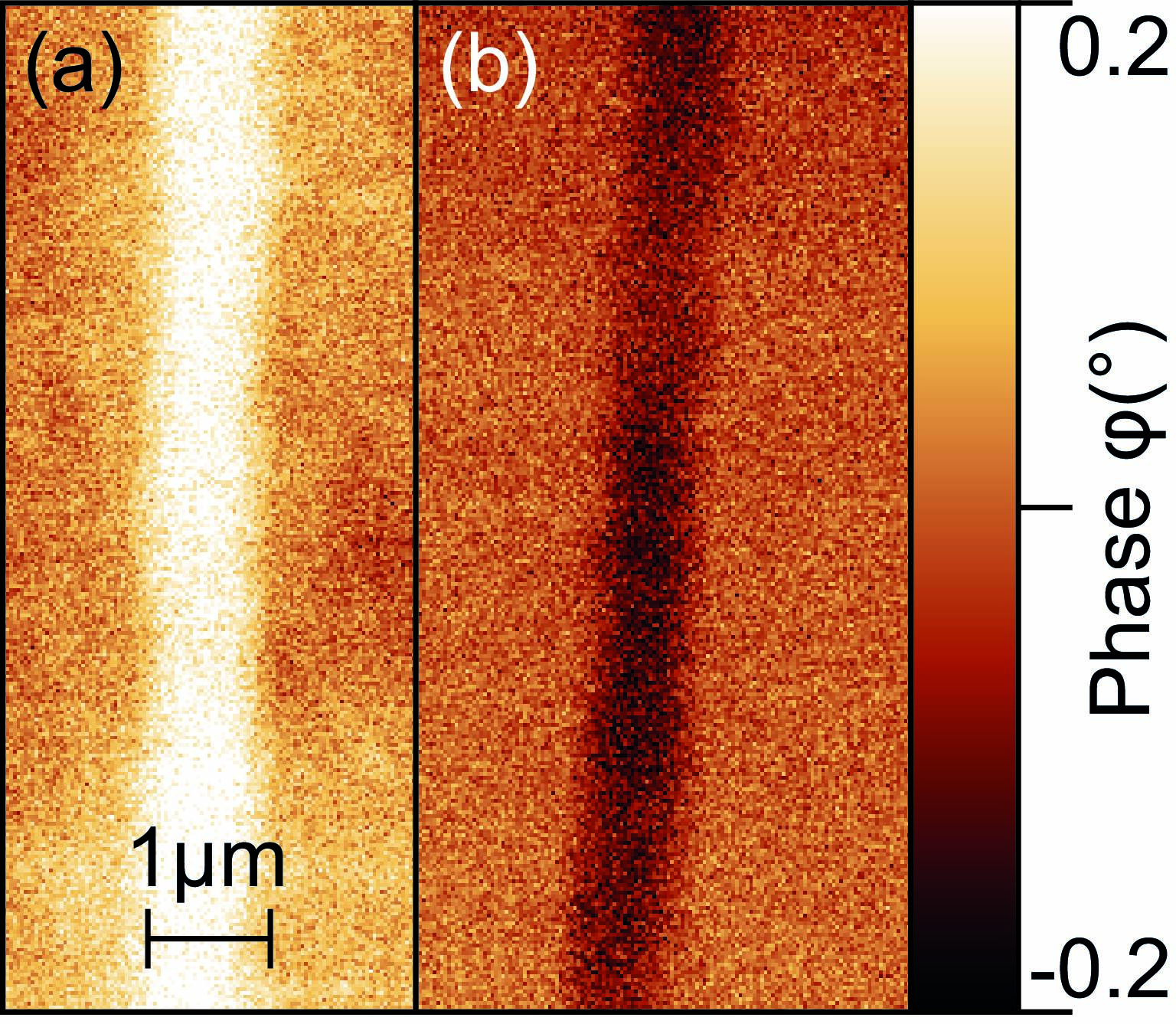}\\
\vspace{+0.7cm}
\includegraphics[width=0.9\columnwidth]{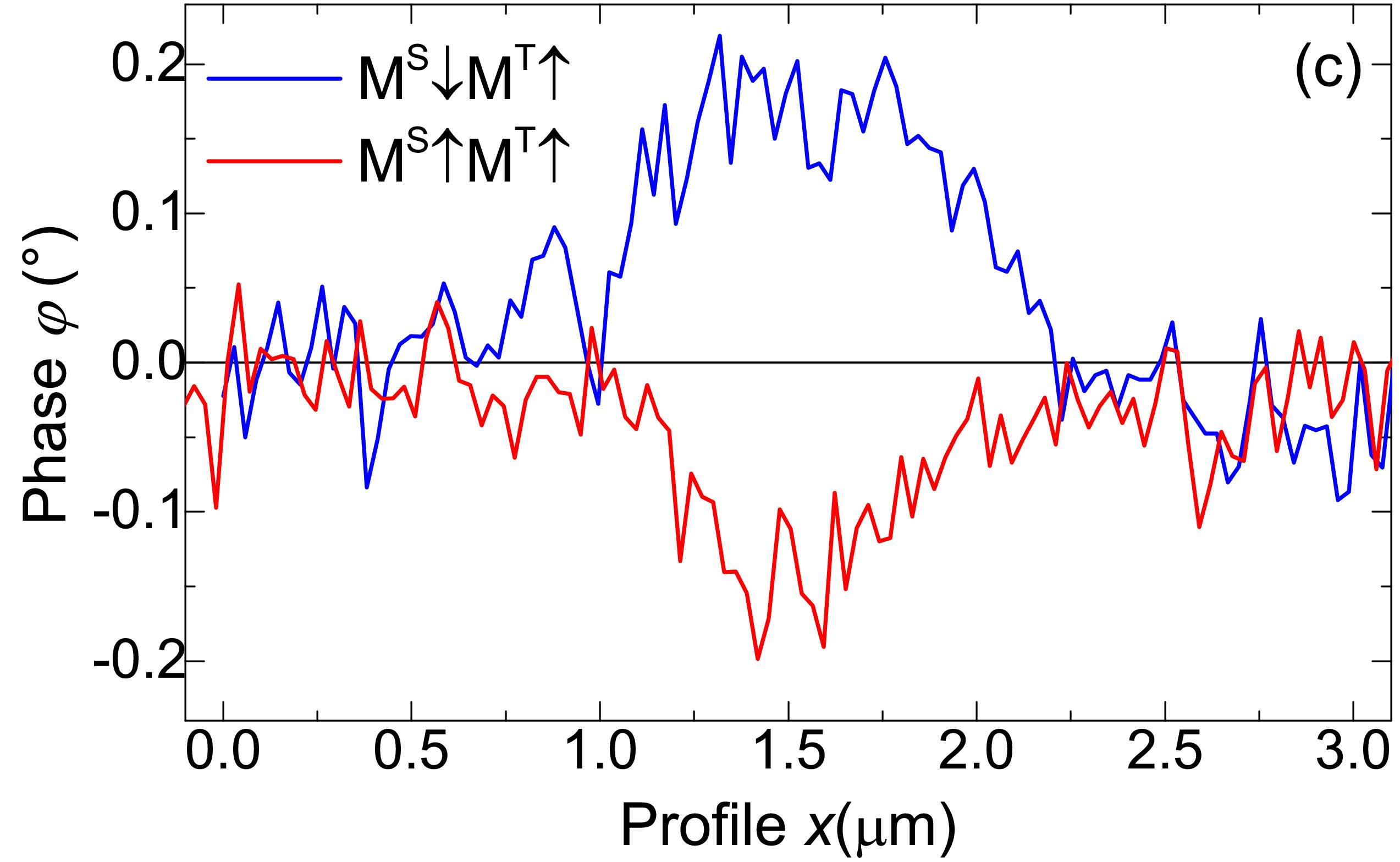}
\end{center}
\caption{Magnetic force microscopy measurements with the sample
  (M$^{\rm S}$) and tip (M$^{\rm T}$) magnetization antiparallel $(a)$ and
  parallel $(b)$ with respect to each other; $(c)$ shows the
  corresponding line scans.}\label{mfm2}
\end{figure}

In order to examine the possibility for controlled magnetic
manipulation at the surface of the anatase thin film, the samples were
patterned using electron beam lithography. Therefore, a film was
covered with a resist, and electron beam lithography was used to
prepare a mask. The resulting irradiated lines or stripes have a width of
$\approx750$~nm. After irradiation with low energy argon ions, the
whole mask was completely removed and the sample was magnetized using
a permanent magnet with a magnetic field aligned perpendicular to the
sample surface and two magnetization directions; parallel and
antiparallel to the tip magnetization. No external field was applied
during the measurement\cite{Stiller2020}. The results are shown in
Figures~\ref{mfm2}(a) and (b), and present a clear MFM signal
corresponding to the two out-of-plane magnetic field directions; the
area of the thin film which was not irradiated does not show any MFM
response. Phase line scans normal to the main length of the FM stripes shown  in Figs.~\ref{mfm2}(a) and (b),
are shown in Fig.~\ref{mfm2}(c). 

\section{Conclusion}

In conclusion, it has been shown that ferromagnetism at room
temperature with perpendicular magnetic anisotropy can be induced in
anatase after irradiating the sample with low-energy ions.  The used
method is remarkably simple and cheap compared to other experimental
methods to produce perpendicular magnetic anisotropy, such as magnetic
heterostructures~\cite{li_2d_2021}.  \textcolor{black}{The irradiation strategy is 
similar to the doping approach used in the semiconductor industry. However, 
the advantage of our method relays in its efficiency and the possibility to
easily combine with other techniques, as electron beam
lithography, allowing the production of arbitrary magnetic patterns
with 2D perpendicular magnetic anisotropy.}

\section*{Acknowledgements}
The authors acknowledge support from the German Research Foundation (DFG) and Universit\"at Leipzig within the program of Open Access Publishing

\bibliographystyle{Frontiers-Vancouver} 

\section*{Funding}
Part of this study was supported by the DFG, Project No. 31047526, SFB 762 “Functionality of oxide interfaces,” project B1.
\section*{Acknowledgments}
The authors thank  L. Botsch and W. Hergert  for the discussions, cooperation and support. 
\end{document}